\documentclass[aps,prl, reprint, letterpaper,longbibliography]{revtex4-1}

\usepackage{amssymb}
\usepackage{amsmath}
\usepackage{graphicx}
\usepackage{extarrows}
\usepackage{url}
\usepackage{varioref}
\usepackage{hyperref}
\usepackage{verbatim}

\begin{document}

\title{Super-resolution microscopy of cold atoms in an optical lattice}
\author{Mickey McDonald}
\thanks{These authors contributed equally to this work.}
\affiliation{James Franck Institute, Enrico Fermi Institute and Department of Physics, University of Chicago, Chicago, Illinois 60637, USA}
\author{Jonathan Trisnadi}
\thanks{These authors contributed equally to this work.}
\affiliation{James Franck Institute, Enrico Fermi Institute and Department of Physics, University of Chicago, Chicago, Illinois 60637, USA}
\author{Kai-Xuan Yao}
\affiliation{James Franck Institute, Enrico Fermi Institute and Department of Physics, University of Chicago, Chicago, Illinois 60637, USA}
\author{Cheng Chin}
\thanks{Corresponding author: cchin@uchicago.edu}
\affiliation{James Franck Institute, Enrico Fermi Institute and Department of Physics, University of Chicago, Chicago, Illinois 60637, USA}

\begin{abstract}
Super-resolution microscopy has revolutionized the fields of chemistry and biology by resolving features at the molecular level.
Such techniques can be either ``stochastic'' \cite{betzig2006imaging}, gaining resolution through precise localization of point source emitters, or ``deterministic'' \cite{hell1994breaking, gustafsson2005nonlinear}, leveraging the nonlinear optical response of a sample to improve resolution.
In atomic physics, deterministic methods can be applied to reveal the atomic wavefunction and to perform quantum control.
Here we demonstrate super-resolution imaging based on nonlinear response of atoms to an optical pumping pulse.
With this technique the atomic density distribution can be resolved with a point spread function FWHM of 32(4) nm and a localization precision below 1~nm.
The short optical pumping pulse of 1.4~$\mu$s enables us to resolve fast atomic dynamics within a single lattice site.
A byproduct of our scheme is the emergence of moir\'{e} patterns on the atomic cloud, which we show to be immensely magnified images of the atomic density in the lattice.
Our work represents a general approach to accessing the physics of cold atoms at the nanometer scale, and can be extended to higher dimensional lattices and bulk systems for a variety of atomic and molecular species.
\end{abstract}

\maketitle

\begin{figure}[ht]
\centering
\includegraphics[width=85mm]{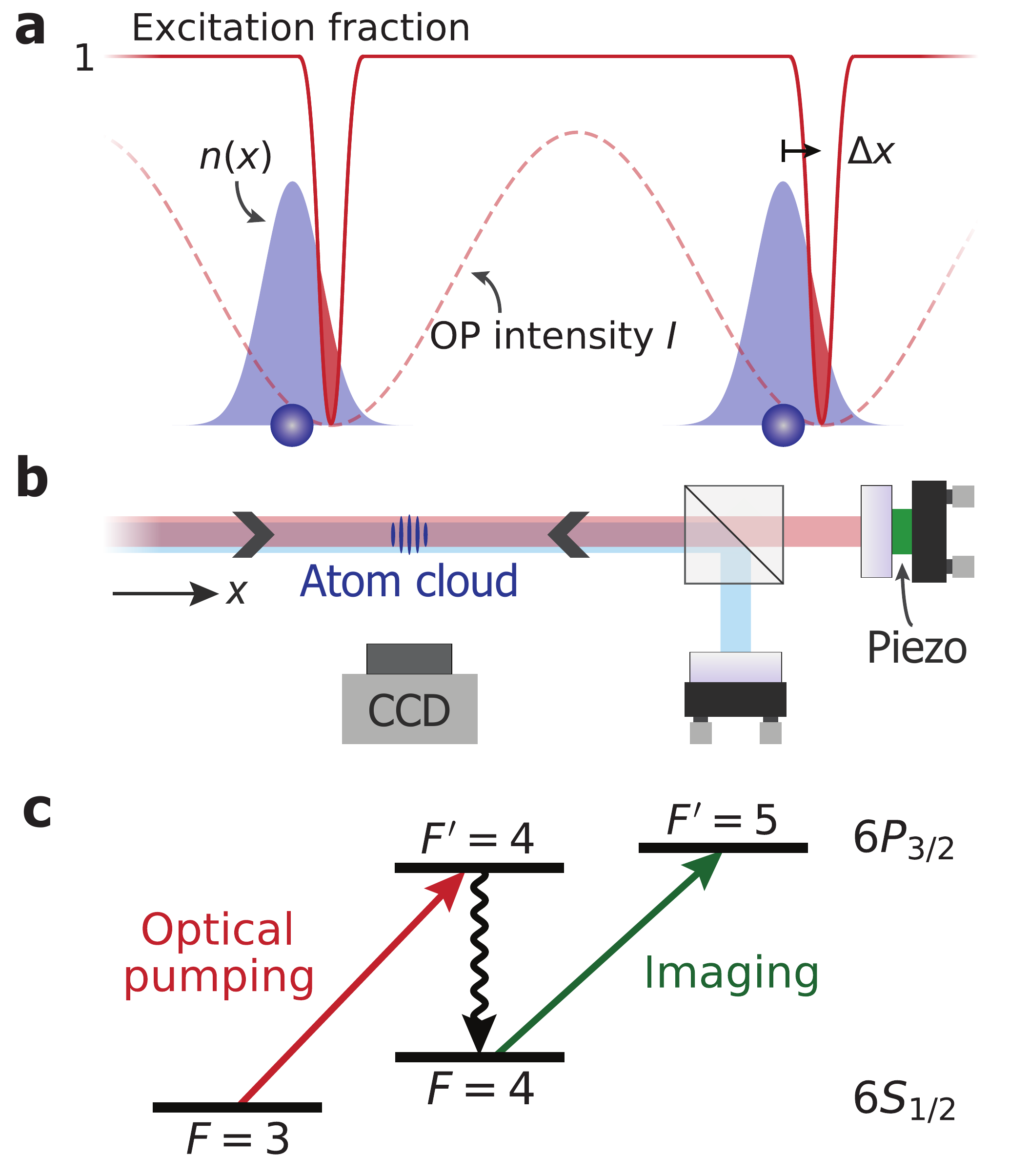}
\caption{\textbf{Super-resolution imaging of ultracold atoms based on optical pumping.}
\textbf{a,} Given an intense standing wave of optical pumping (OP) light, the excitation probability (red solid line) is nearly unity unless the atoms (blue shaded area) are within a narrow window at the nodes of the OP lattice (red dashed line) where the intensity vanishes.
Red shaded area approximates the fraction of the atoms that are not excited.
\textbf{b,} The trapping lattice (cyan) and OP lattice (red) overlap on the atoms.
The relative displacement between the two lattices is controlled by a piezo transducer behind one of the retro reflection mirrors.
The atomic distribution is measured by scanning the piezo displacement  $\Delta x$.
A CCD camera from the side images atoms \textit{in situ} on the strong $|F=4\rangle\rightarrow |F'=5\rangle$ cycling transition.
\textbf{c,} Relevant atomic levels.
Atoms initially in the $|F=3\rangle$ ground state are optically pumped to $|F=4\rangle$, and imaged with the imaging beam.
Primed and unprimed letters refer to levels within the excited state $6^2P_{3/2}$ and ground state $6^2S_{1/2}$ manifolds, respectively.
}
\label{figSchematic}
\end{figure}

In the study of ultracold atomic gases, high resolution microscopy has played an important role in visualizing interesting quantum phenomena. Examples include phase transitions~\cite{gemelke2009situ,parker2013direct}, correlations~\cite{endres2011observation,cheneau2012light,hung2013from}, transport~\cite{brantut2012conduction}, tunneling ~\cite{kaufman2014two}, and quantum information processing with ions~\cite{blatt2008entangled} and atoms~\cite{nelson2007imaging,saffman2010quantum}.
Optical microscopy of ultracold gases has been pushed to its limit to detect atoms in optical lattices with sub-micron spacings~\cite{bakr2009quantum,sherson2010single,omran2015microscopic,cheuk2015quantum,parsons2015site,edge2015imaging,yamamoto2016ytterbium}.
The spatial resolution in these experiments is constrained by the imaging wavelength to typically $0.5\sim1$~$\mu$m, a value set by the Abbe limit $d=\lambda/2\mathrm{NA}$ \cite{mccutchen1967superresolution}.
Here, $\lambda$ is the wavelength of the imaging light and $\mathrm{NA}$ is the numerical aperture of the microscope.

Several schemes have been demonstrated which reach beyond the optical diffraction limit.
Scanning electron microscopy of ultracold gases visualizes atoms with a resolution of 150 nm~\cite{gericke2008high}.
Stochastic techniques are applied to localize the mean  positions of trapped ions to a few nanometers~\cite{wong2016high}, as well as the occupancy of closely-spaced one-dimensional (1D) optical lattice sites~\cite{alberti2016super}.
Stochastic methods, however, derive their power from the assumption of point-source emission, meaning that the atomic wavefunction itself cannot be resolved.

Another class of deterministic super-resolution imaging with genuine sub-wavelength resolution exploits the nonlinear optical response of atoms to a spatially varying light field.
Proposals exist which are based on spatially dependent coherent dark state transfer~\cite{gorshkov2008coherent,cho2007addressing,agarwal2006subwavelength,paspalakis2001localizing}.
These proposals hold the promise to resolve atomic wavefunctions and their dynamics in an optical lattice.

In this paper we demonstrate 1D super-resolution microscopy of ultracold atoms at the nanometer scale.
Our technique shares conceptual similarities to saturated structured illumination microscopy (SSIM) \cite{gustafsson2005nonlinear} and stimulated emission depletion (STED) microscopy \cite{hell1994breaking}, and is schematically illustrated in Fig.~1.
Atoms are initially localized in the trapping lattice and polarized in the ${|F=3\rangle}$ ground state, where $F$ is the total angular momentum.
An additional optical pumping (OP) lattice is applied which pumps atoms to a different hyperfine state ${|F=4\rangle}$.
Since just a few photons are required to pump atoms to the new state, only atoms within a narrow window around the nodes of the OP lattice are likely to remain in the initial state, while those outside of this window have near-unity probability to be pumped to the ${|F=4\rangle}$ state.
By sweeping the location of this window across the atomic density distribution and measuring the fraction of atoms remaining in ${|F=3\rangle}$, a map of the atomic density distribution can be built up with a resolution given by the width of the window.
As we will discuss below, this width can be made arbitrarily small compared to the optical wavelength, which is key to attaining high resolution.

Our experimental implementation is illustrated in Fig.~1b.
A cloud of $^{133}$Cs atoms is collected in a magneto optical trap, subsequently cooled by degenerate Raman-sideband cooling to $<1\mu$K \cite{kerman2000beyond}, and polarized in the ${|F=3\rangle}$ ground state.
About $2\times 10^6$ atoms are then adiabatically loaded into a one-dimensional optical lattice with approximately 90\% occupation in the motional ground state along the lattice direction.
The trapping lattice with lattice constant $\lambda_\mathrm{trap}/2 \approx 426$~nm is blue-detuned from the resonance transition ${|F=3\rangle}\rightarrow {|F'=4\rangle}$ by $\delta=+10$ to $500$~GHz.
The OP laser is resonant with the ${|F=3\rangle}\rightarrow {|F'=4\rangle}$ transition at $\lambda_{\mathrm{OP}}=852.335$~nm (see Fig.~1c), and is retro-reflected and polarized perpendicularly to the co-propagating trapping lattice.
The retro-reflection of the OP beam is carefully aligned and balanced to cancel the electric field at the nodes of the standing wave.
The relative phases of the two lattices are controlled with nanometer precision using a piezoelectric transducer (see Methods).

To image the atoms, we apply a 1.4~$\mu$s pulse of the OP lattice,
which transfers atoms to ${|F=4\rangle}$.
This pulse is short compared to the timescale of atomic motion, but much longer than the $6P_{3/2}$ excited state lifetime of $1/\Gamma=30$~ns.
The OP pulse is followed by \textit{in situ} imaging with a camera in the direction perpendicular to the lattice.
From measuring the atomic population in ${|F=4\rangle}$, we determine the excitation fraction $\mathcal{F}$ across the sample.

\begin{figure*}[htp]
\centering
\includegraphics[width=170mm]{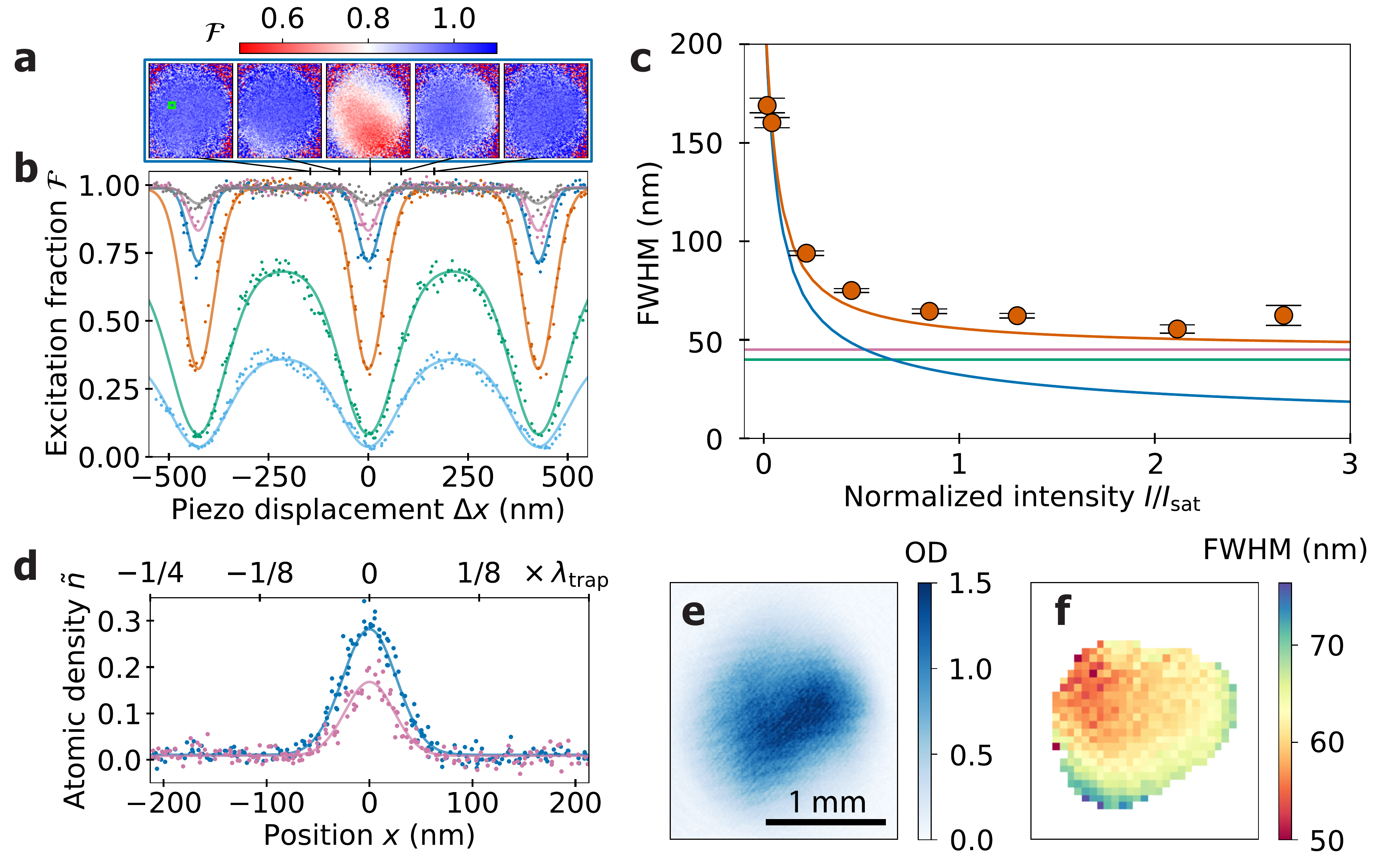}
\caption{\textbf{Performance of super-resolution imaging.} 
\textbf{a,} Images of excitation fraction $\mathcal{F}$ taken at different piezo displacements $\Delta x$.
For these images, the OP intensity is $I/I_\mathrm{sat}=1.3$, and atoms are prepared in a trapping lattice with detuning $\delta=10~$GHz and lattice depth $U=200\mu$K.
The region outlined in green indicates the area from which the data in panels \textbf{b}-\textbf{d} are collected.
\textbf{b,} As the optical pumping intensity increases, the atom response becomes more nonlinear and the dips in the excitation fraction narrow.
From low to high, the traces show measurements at increasing intensity with $I/I_\mathrm{sat}= 0.017 (\textrm{light blue}), 0.041 (\textrm{green}), 0.22 (\textrm{orange}), 1.3 (\textrm{blue}), 2.1 (\textrm{purple}), 2.7 (\textrm{grey})$.
Solid curves show fits based on a sum of Gaussians separated by $\lambda_\mathrm{OP}/2$. 
\textbf{c,} FWHMs of the fitted Gaussians at different OP intensities (orange circles) are compared with those of the ground state (40~nm, green), a thermal state with $90\%$ ground state occupation (45~nm, purple), theoretical resolution (blue) and the expected width (orange) of the convolution of the thermal state and the point spread function $g(x)$ (see Supplemental Information). Error bars show one standard error.
\textbf{d,} Derived single site atomic density distribution.
The measurement reflects the density distribution averaged over sites contained within the green-outlined region in panel \textbf{a}.
From Gaussian fits, we determine FWHM=55(2) and 62(1)~nm, and uncertainty in the peak positions of 0.8 and 0.4~nm, for $I/I_\mathrm{sat}=2.1 (\textrm{purple})$ and $1.3 (\textrm{blue})$, respectively.
\textbf{e, }A typical optical density (OD) image of all atoms.
\textbf{f, }Distribution of the FWHM across the cloud at $I/I_\mathrm{sat}=2.1$.
Only area containing signal sufficient for fitting is shown.
}
\label{figResolution}
\end{figure*}

To explore the resolving power of this technique, we record traces of the excitation fraction $\mathcal{F}$ versus piezo displacement $\Delta x$ (see Fig.~2).
At sufficiently low OP beam intensities $I\ll I_\mathrm{sat}$ the excitation fraction $\mathcal{F}(\Delta x)$ varies sinusoidally, mirroring the sinusoidal intensity profile of the OP lattice $I(x)=4I\sin^2(2\pi x/\lambda_{\mathrm{op}})$.
At higher intensities, however, the response of atoms to optical pumping becomes more nonlinear because the excitation fraction quickly saturates to 1 unless atoms are located sufficiently close to the nodes of the OP lattice.
In this regime the remaining fraction $g(x)$ of atoms in the ${|F=3\rangle}$ state near a node is approximately given by
\begin{equation}
    g(x) = \exp\left[-\frac{\beta \Gamma }{2}\frac{I(x)}{I(x)+I_\mathrm{sat}}t\right],
    \label{eqNonLin}
\end{equation}
where we have assumed a long exposure time $t\gg1/\Gamma$ and  $\beta=7/12$ is the branching ratio of spontaneous emission into the ${|F=4\rangle}$ state. At the nodes, $g(x)$ develops narrow peaks.

The narrowing of the excitation dips at higher OP intensity (see Fig.~2b) results from the nonlinear optical response described in Eq.~(\ref{eqNonLin}).
This narrowing can also be understood as revealing the atomic density distribution with increasing resolving power.
Given a spatial density distribution $n(x)$ for an atom (in either a pure or mixed quantum state) under the spatially varying OP intensity $I(x)$, the excitation fraction $\mathcal{F}(\Delta x)$ directly relates to the atomic density $n(x)$ as
\begin{equation}
1-\mathcal{F}(\Delta x) = \int n(x) g\left(\Delta x-x\right) dx \equiv \tilde{n}(\Delta x),
\label{eqConv}
\end{equation}
where $\tilde{n}(\Delta x)$ is the convolution of the atomic density distribution with the point spread function given by $g(x)$.
When the width of $g(x)$ is smaller than that of the atomic density distribution, $\tilde{n}(\Delta x)$ (and, equivalently, $1-\mathcal{F}$) reveals the atomic density distribution (see Figs.~2c and 2d).
Because the excitation fraction $\mathcal{F}$ is measured with a finite imaging resolution, the extracted density distribution $\tilde{n}(\Delta x)$ is an average over sites contained in the resolution limited spot.

For an OP pulse of duration $t\gg 1/\Gamma$, the imaging resolution is defined based on the full width at half maximum (FWHM) $w$ of the point spread function $g(x)$, and is calculated to be (see Supplementary Information)
\begin{equation}\label{eqresolution}
    w = \frac{\lambda_\mathrm{op}}{2\pi}\sqrt{\frac{2\ln{2}}{s t \beta\Gamma}},
\end{equation}
where $s = I/I_{\mathrm{sat}}$ and $I$ is the single beam intensity. 
In our experiment, the calculated imaging resolution above $s=0.6$ is high enough to reveal the shape of our atomic density distribution (see Fig. 2c).

Our measured widths, reaching a minimum of $55(2)$~nm, are in good agreement with the expected widths from the theory prediction (see Figs.~2c and 2d). 
From the measurement at $s=2.1$ we calculate an imaging resolution, defined by the FWHM of the extracted point spread function \cite{hell1994breaking}, to be $32(4)$~nm, which is less than $1/25$ of the $852$~nm imaging wavelength.
Furthermore, from Gaussian fits, the center positions of the atomic density can be localized to about $0.4$~nm.
Notably, the imaging resolution worsens at very high OP intensity $s>2.5$ because of the limited signal-to-noise ratio.

Applying the same analysis everywhere in the image allows us to map the fitted widths across the cloud (see Fig.~2f). A variation of 40\% is seen, likely due to the combination of inhomogenous cooling efficiency and trap depth. We note that such spatially-resolved information about trap parameters is often inaccessible  using conventional imaging techniques. 

\begin{figure*}[htp]
\centering
\includegraphics[width=170mm]{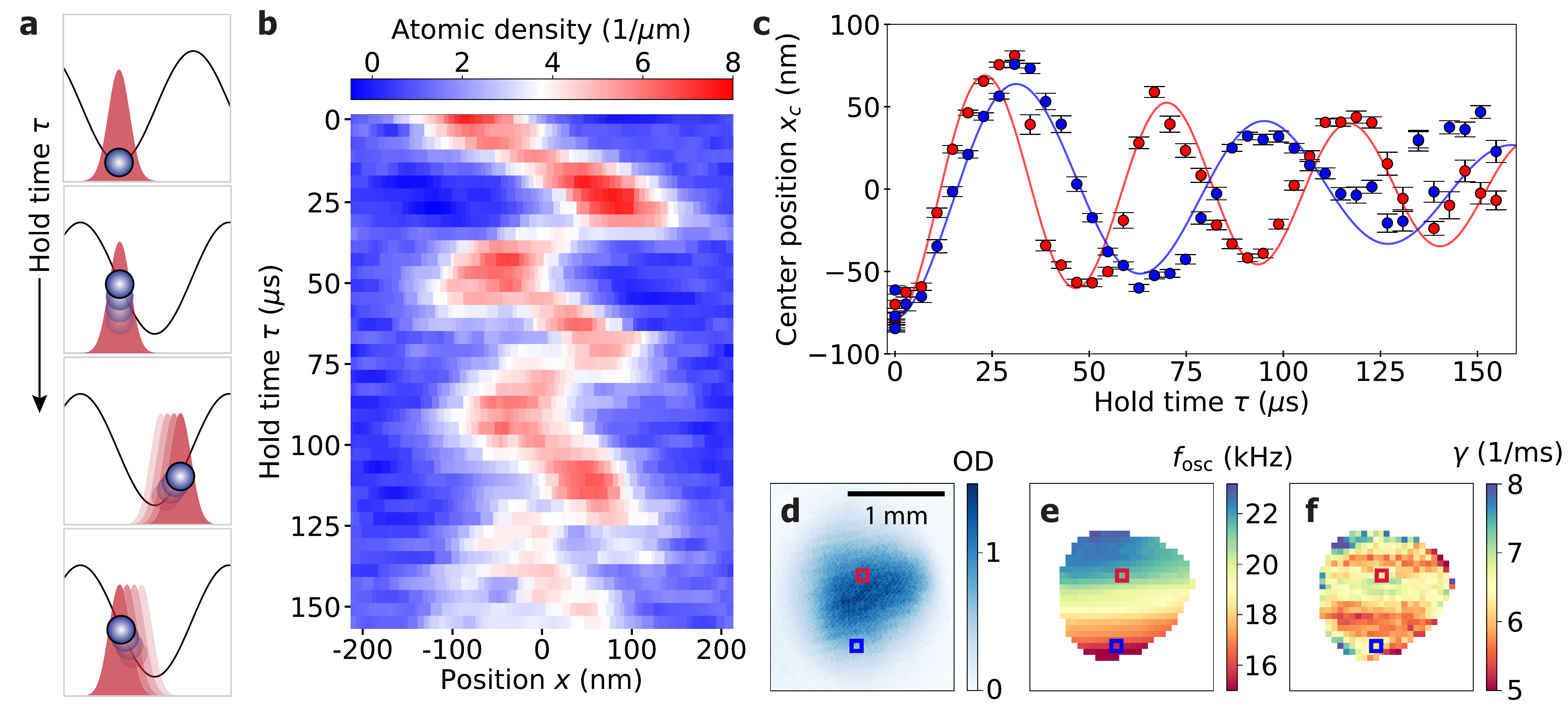}
\caption{\textbf{Microscopic dynamics revealed by super-resolution microscopy.}
\textbf{a,} A cartoon illustrating the experimental procedure.
After preparing atoms in the ground state, the trapping lattice is displaced by an amplitude $A=79$~nm, causing the atoms to oscillate.
\textbf{b,} Atomic density evolution within a single lattice site.
After a hold time $\tau$ we record the atomic density using the same procedure as Fig.~2d.
Here the trapping lattice detuning is $\delta = 20$~GHz, lattice depth $U =10$~$\mu$K and OP intensity $I/I_\mathrm{sat}=1.3$.
Jagged motion of the atoms results from anharmonicity of the lattice potential (see Supplementary Information).
Data are extracted from the location shown in the blue box in panel \textbf{d}.
\textbf{c,} Evolution of the atomic position determined from Gaussian fits.
Blue and red circles are based on measured data from bins of the same color in panel \textbf{d}.
Solid curves show exponentially decaying cosine fits $x_c = -Ae^{-\gamma\tau}\cos{2\pi f_\mathrm{osc}\tau}$, where $\gamma$ is the decay constant and $f_\mathrm{osc}$ is the oscillation frequency.
Error bars show one standard error.
\textbf{d,} Typical absorption image of the atomic cloud.
\textbf{e,} Map of fitted oscillation frequency $f_\mathrm{osc}$ across the cloud.
\textbf{f,} Map of fitted decay constant $\gamma$ across the cloud.}
\label{figDynamics}
\end{figure*}

An important feature of this imaging scheme is the short $\mu$s duration of the OP pulse compared to the timescale of typical atomic motion in the lattice. Our scheme is thus ideally suited for probing dynamics of atoms within a lattice site.
To explore this capability, we quickly displace the trapping lattice by 79~nm and record the evolution of the atomic density distribution after different hold times (see Fig.~3).
 
The displacement initiates an oscillatory motion of the atoms (see Fig.~3b). The ``jagged'' features of the motion come from the anhamonicity of the lattice potential (see Supplementary Information). From the time evolution, we further extract the oscillation frequency and damping rate of the atomic motion, as shown in Fig.~3c for two bins in separate locations, and construct the complete maps of these quantities in the sample (see Figs.~3d and 3e), which clearly show the inhomogeneity of lattice parameters.

\begin{figure*}[htp]
\centering
\includegraphics[width=170mm]{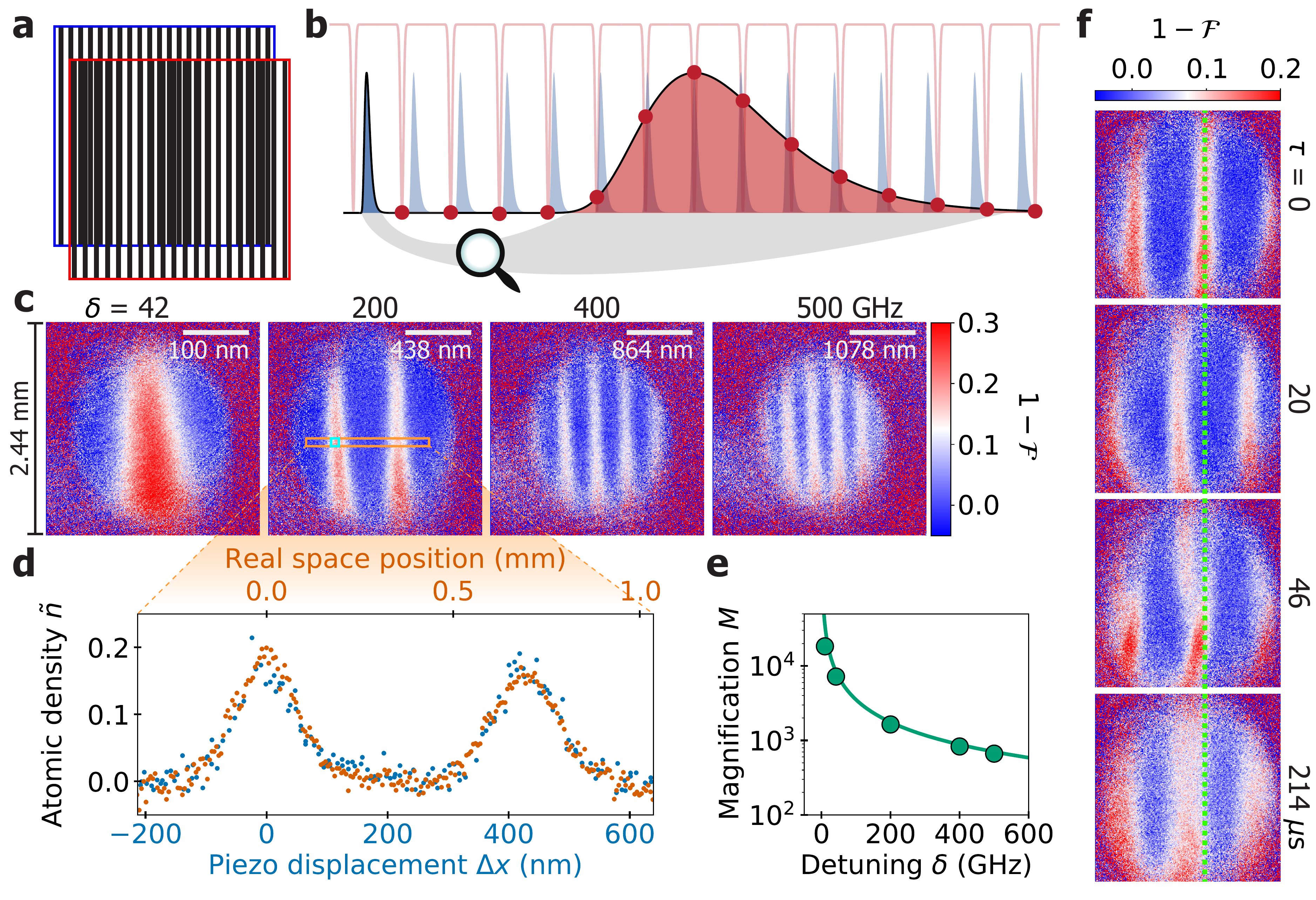}
\caption{\textbf{Moir\'{e} magnification of the atomic density distribution.} 
\textbf{a,} When two periodic structures of slightly different pitch are overlapped, a large-scale moir\'{e} pattern emerges.
\textbf{b,} Magnification of the atomic density distribution based on moir\'{e} interference. 
The OP lattice (red) interrogates the atomic density distribution (blue shaded) at different positions (red dots) within each confining lattice site.
If the atomic density distribution in every site is identical, then the resulting excitation fraction across the cloud traces out the density distribution with large magnification.
\textbf{c,} \textit{In situ} images of excitation fraction taken at different lattice detunings $\delta$ with a field of view of (2.44~mm)$^2$ with $I/I_\mathrm{sat}=0.89$.
In each image the moir\'{e} pattern reflects the microscopic atomic density distribution.
Spacing between stripes is $\lambda_\mathrm{trap}/2$.
White bars show the microscopic length scale.
\textbf{d,} Comparison of the moir\'{e} pattern within the orange rectangle in panel \textbf{c} (orange circles) to the microscopic atomic density distribution measured in the cyan box in panel \textbf{c} by piezo scan (cyan circles). 
The horizontal axis of each data set is scaled to match spatial periodicity, and translated to overlap the peaks.
\textbf{e,} Dependence of moir\'{e} magnification $M$ on detuning $\delta$.
The solid line is based on Eq.~(\ref{eqMag}).
See Supplementary Information for details.
\textbf{f,} Evolution of the moir\'{e} pattern after a 79~nm lattice phase jump with detuning $\delta = 200$~GHz.
After hold time $\tau$ the moir\'{e} pattern oscillates and becomes distorted, indicating that the dynamics of the atoms in the lattice are not uniform across the sample.
The dashed green line serves as a reference indicating the initial position of the center stripe.
}
\label{figMoire}
\end{figure*}

Thus far all measurements have required repeating the experiment many times, each with a small increment in the piezo displacement. Here we develop an alternative method that exploits the slight difference in wavelengths of the optical pumping and trapping beams to obtain the atomic density distribution at nanometer scale in a single shot based on the moir\'{e} effect.

When two gratings of slightly different periodicity overlap, a moir\'{e} interference pattern emerges at a macroscopic length scale (see Fig.~4a for an example) because the relative phase of the two gratings advances slowly and linearly along the grating direction.

In our experiment, the slight difference in the wavelengths of the trapping lattice $\lambda_{\mathrm{trap}}$ and OP lattice $\lambda_\mathrm{OP}$ causes the atoms trapped in neighboring lattice sites to be probed at slightly different positions within each site.
If the atomic density profile is identical along the lattice direction, the resulting moir\'{e} pattern imprinted onto the cloud represents a greatly magnified image of the density profile (see Fig.~4b).
The magnification $M$ is given by~\cite{morse1960geometry} 
\begin{equation}
    M = \frac{\lambda_\mathrm{OP}}{|\lambda_\mathrm{OP}-\lambda_\mathrm{trap}|},
    \label{eqMag}
\end{equation}
which in our experiment can expand 10~nm features to 1~mm scale in a single shot image.

Figure 4c shows a representative series of moir\'{e} patterns of excitation fractions observed at different detunings of the trapping lattice. The stripes, appearing with greater number at larger detuning, show the rephasing of the two lattices, and the separation between two stripes corresponds to the microscopic lattice constant.  

To confirm that the moir\'{e} patterns represent a faithful magnification of the atomic density distribution in a lattice site, we compare the pattern to the density profile extracted from piezo scanning (see Fig.~4d).
Here a weaker lattice is chosen so that the measured width is dominated by that of the atoms. 
The two measurements match excellently, which confirms the interpretation of a moir\'{e} pattern as a magnified image of atomic density distribution in a lattice site. We determine the magnification for each image in Fig.~4c and the result also shows good agreement with Eq.~(4). At the smallest detuning of $10~$GHz in our experiment, the magnification reaches $M=20,000$. 
 
This moir\'{e} pattern based imaging scheme is also a convenient tool to study the atomic dynamics in the lattice.
After displacing the lattice by 79~nm, the moir\'{e} pattern appears straight in the beginning, but develops snaking wiggles after 20$\sim$30 $\mu$s and finally relaxes to a wider stripe at a displaced location (see Fig.~4f).
The snaking wiggles in each stripe indicates the inhomogeneous trap parameters across the cloud, confirming the observation in Fig.~3c based on piezo tuning. 

In summary, we demonstrate a super-resolution imaging scheme for cold atoms, which achieves spatial resolution of 32(4)~nm and localization of $<1$~nm by exploiting the nonlinear response of atoms to optical pumping.
The method is ideal for probing the atomic wavefunction in a lattice site.
In addition, the short $\mu$s pumping time allows for resolving fast atomic dynamics. 
For an array of atoms with identical wavefunctions, we also show that moir\'{e} interference patterns can offer macroscopic views of the wavefunction with magnification reaching $20,000$. 

Our imaging method is generic and can be readily applied to other atoms and molecules.
Extension of the method to two and three dimensions is straightforward.
By implementing this scheme in a system with single-site imaging resolution (e.g. quantum gas microscopes), one can gain full information of the quantum system at every site, down to the nanometer scale.
\\
M. M. is supported by the Univ. of Chicago PSD Dean's postdoc fellowship. We thank G. Downs, P. Ocola, and M. Usatyuk for assistance in the construction of the experiment, and L. W. Clark for carefully reading the manuscript. This work is supported by the Army Research Office under Grant No. W911NF-15-1-0113, and the University of Chicago
Materials Research Science and Engineering Center, which
is funded by the National Science Foundation under Grant
No. DMR-1420709.

\section*{Methods}

\noindent\textbf{Optical setup.} To take full advantage of the nonlinear optical response described by Eq.~(\ref{eqNonLin}), it is critical that the OP lattice has clean zero-intensity nodes. 
Due to small losses accumulated in the optical path (e.g. from windows, beamsplitters, etc.), the retro-reflecting beam diameter is made to be 84\% the incident beam diameter so that incident and retro intensities can be closely matched.
Additional fine tuning of the retro intensity is provided by adjusting its transmission through a polarizing beam splitter using a quarter waveplate. 
The retro intensity is optimized by maximizing the signal-to-noise of $\tilde{n}$ at $I/I_\mathrm{sat}\gtrsim 1$.

Precise alignment of the OP and trapping lattices is necessary to minimize blurring due to angled moir\'{e} fringes (see Supplementary Information).
We do so by outputting both beams from the same optical fiber, and precisely aligning their retro-reflections via fiber back-coupling to within $\pm$ 20~$\mu$rad of optimal.
This procedure is performed within a few hours before experiments are run to correct for mirror drift.

\noindent\textbf{Preparation of atoms in a 1D optical lattice.} 
We prepare the sample by performing degenerate Raman sideband cooling in a 3D lattice with trap frequency $\omega_\mathrm{trap}\sim 2\pi \times 30$~kHz (measured via phase modulation of the lattice) for 40~ms. 
After cooling, the atoms are polarized in the ${|F=3\rangle}$ hyperfine state and are then adiabatically loaded in 1~ms into a 1D trapping lattice. 
Through time-of-flight temperature measurements, we determine 90\% occupancy in the motional ground state in the lattice direction.


\noindent\textbf{Dynamics.} For the dynamics experiment described in Fig.~3, the 1D trapping lattice is translated by jumping the laser frequency by 56~MHz in $\sim$3~$\mu$s using an acousto-optic modulator (AOM).
This corresponds to a positional shift of 79~nm of the lattice sites given a separation of 0.50 m between the atom cloud and the retro mirrors.
After a variable hold time $\tau$, the atomic density is sampled with the OP pulse as described above. 
The data presented in Fig.~3b are smoothed using a local low-order regression with a window of $\lambda_\mathrm{trap}/20$ at each hold time $\tau$.
Note that while the entire imaging sequence spans 10~$\mu$s, the relevant signal is accumulated only during the 1.4~$\mu$s OP pulse, which allows for studies of fast $<$~10~$\mu$s dynamics as described in the main text.

\noindent\textbf{Moir\'{e} magnification.} When imaging moir\'{e} patterns, the experimental procedure is identical to the generic super-resolution experiment, except that the retro mirror displacement is not varied.
Post-processing to obtain $\mathcal{F}$ is identical, except no binning is used.
The size of the camera pixels in the imaging plane is determined by dropping the cloud and comparing its acceleration to gravity.
The moir\'{e} magnification $M$ in Fig.~4e is obtained by dividing the real space distance between stripe centers (as fitted by Gaussians) by the lattice spacing $\lambda_\mathrm{trap}/2$.
Moir\'{e}-magnified dynamics are realized by performing the same lattice phase jump as described in Fig.~3.

\noindent\textbf{Post-processing.} Post-processing consists of binning each image (typically using 10-pixel wide squares).
Bins with small atom number below a threshold have low signal-to-noise ratio and are therefore not analyzed.
Due to a frequency difference between OP and trapping lattices, this 10-pixel binning contributes a small (few~nm) blur in the signal (see Supplementary Information).
The size of 10 pixels is chosen to balance between good signal-to-noise and blurring.
We obtain images of atoms with the OP lattice and a reference image with all atoms pumped to ${|F=4\rangle}$ in order to determine the excitation fraction $\mathcal{F}$ for each bin.

\setcounter{figure}{0} 
\setcounter{equation}{0}

\renewcommand{\theequation}{S\arabic{equation}}
\renewcommand{\thefigure}{S\arabic{figure}}

\setcounter{page}{1}

\section*{Supplemental Information for: \\ Super-resolution microscopy of cold atoms in an optical lattice}
\subsection{Description of experiment}

\subsubsection*{Preparation of cold atoms in an optical lattice}
The experiment begins by loading a magneto-optical trap for 1~s with $\sim2\times 10^7$ $^{133}$Cs atoms and performing molasses cooling to $\sim$10~$\mu$K. After, we turn on a 3D optical lattice (trap frequency $\approx$30~kHz) to perform degenerate Raman sideband cooling down to $<$1 $\mu$K, after which we are left with $\approx 2\times 10^6$ atoms polarized in the ${|F=3,m_F=3\rangle}$ state with 90\% occupancy in the vibrational ground state (as determined by time-of-flight thermometry). After cooling, two axes of the 3D trapping lattice are adiabatically ramped off in 1~ms and the remaining trapping 1D lattice (spacing = $\lambda_\mathrm{trap}/2$) is ramped to a chosen power which determines the single-site harmonic oscillator width. At this point the sample is ready for the super-resolution experiment.

\subsubsection*{Optical setup}
Good beam alignment of the 1D OP lattice and the 1D confining lattice is crucial to achieving high resolution.
To ensure coincidence of the two beams, they are combined in a polarizing beam splitter and then fiber-coupled to a polarization-maintaining fiber (with orthogonal linear polarizations).
Formation of near-zero intensity nodes on the OP lattice is critical to obtaining high SNR.
To account for optical loss accumulated (e.g. at windows, beam-splitters), the fiber output passes through two lenses which weakly focus the beam such that at the atom location, the retro-reflected beam diameter is $\approx$84$\%$ the incident beam diameter so that the intensities are closely matched.
Additionally, fine adjustment of the retro-reflected intensity is provided by a $\lambda/4$ waveplate (QWP2).
The tip and tilt of the retro-reflected beams are each aligned to within $\pm$20 $\mu$rad on a daily basis via precise back-coupling into the fiber.

\begin{figure*}[tb]
\begin{center}
\includegraphics[width=160mm]{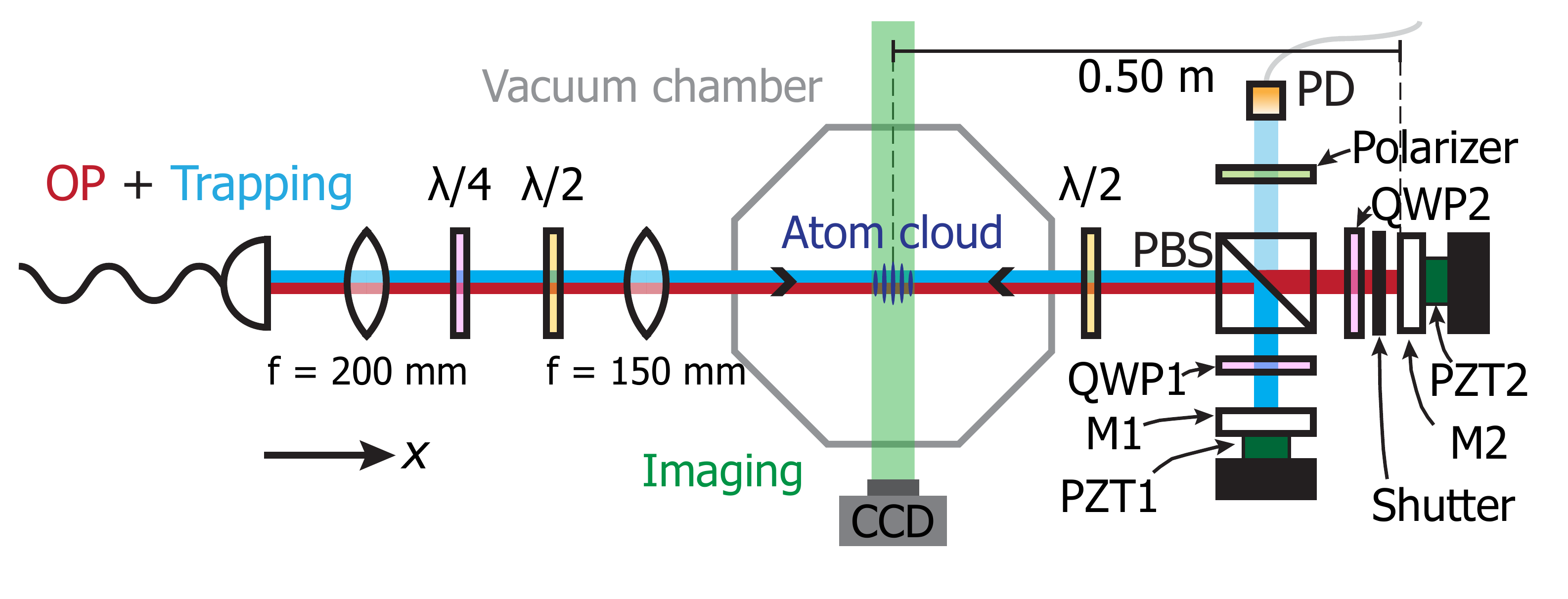}
\caption{\textbf{Optical setup for super-resolution experiment.} The 1D OP and 1D trapping lattices emerge from the same fiber to ensure good relative alignment. A polarizing beam-splitter (PBS) is used to separate the two beams, which are linearly polarized in orthogonal directions. The photodiode PD is used to calibrate the relative positions of the two retro-reflecting mirrors, which are both 0.50~m away from the atom cloud.
}
\label{fignumsupp}
\end{center}
\end{figure*}

\subsubsection*{Calibration of piezo displacement}
Calibration of the piezo displacement $\Delta x$ must be accurate to within nanometers in order to prevent systematic distortion of the signal.
Here $\Delta x$ is primarily determined by the relative positions of the two retro-reflecting mirrors of the OP and trapping lattices, which are measured interferometrically every shot.
To perform this measurement, we turn on the trapping lattice and make use of leakage light from the polarizing beam splitter shown in Fig.~1b. 
The two arms of the polarizing beam splitter re-combine and interfere on a photodiode. 
A second piezo, attached to the trapping lattice mirror, is scanned over a few lattice spacings, and the phase (in nm) of the resulting sinusoidal signal on the photodiode is measured.

Additionally, a correction term to $\Delta x$ is applied in order to account for position drift in the trapping lattice nodes due to frequency drift of the laser. 
The trapping lattice originates from a Ti:sapphire laser that is stable to $\approx$50~MHz/hour. 
The nodes of the lattice will shift by $(\Delta f_\mathrm{tr}/f_\mathrm{tr})L$ where $\Delta f_\mathrm{tr}$ is the change in frequency, $f_\mathrm{tr}=351$~THz is the frequency of the trapping light, and $L$ is the distance between the atom cloud and the retro-reflecting mirror.
For our setup, $L=0.50$~m so that the trapping nodes will shift at a rate of 1.4~nm/MHz.
We calibrate the frequency drift of the trapping laser every shot by observing its peak position on a Fabry-Perot cavity relative to a stable reference laser to within 1~MHz.

\subsubsection*{Atom number drift}

Since our signal is the excitation fraction, slow drift in total atom number is calibrated using a running reference image that is based on saturated atom images that are taken using a high-power OP beam without spatial structure (i.e. no lattice). 
These calibration shots are taken either every shot or every four shots, depending on the type of experiment we perform. 
The calibration shots are binned, and the reference image is calculated by applying a Savitsky-Golay filter to each bin using calibration shots nearby in time.

\subsection{Optical pumping under spatially dependent drive field}

Here we derive Eq.~\ref{eqNonLin} in the main text, stating that the excitation fraction under a spatially dependent drive field is given by a convolution. The derivation is given assuming a pure initial state for the atom. Generalization to mixed states is straight forward.

The atom has spatial and electronic degrees of freedom. Therefore, a state can be written as

\begin{equation}
    |\psi\rangle=\sum\limits_{i,j}\psi_{i,j}|i\rangle_x\otimes|j\rangle_e,
\end{equation}
where $|i\rangle_x$ and $|j\rangle_e$ form a basis in the spatial and electronic subspace, respectively, and $\psi_{i,j}$ are the probability amplitudes. 

A density matrix $\hat\rho$ can be written similarly:
\begin{equation}
\begin{split}
    \hat\rho&=\sum\limits_{i_1,i_2,j_1,j_2}\rho_{i_1,i_2,j_1,j_2}|i_1\rangle_x\langle i_2|_x\otimes|j_1\rangle_e\langle j_2|_e\\
    &=\sum\limits_{\alpha,\beta}\rho_{\alpha,\beta}\hat\alpha_x\otimes\hat\beta_e.
\end{split}
\end{equation}
Here each $\hat\alpha_x = |i_1\rangle_x\langle i_2|_x$ for some $i_1,i_2$, and notates a basis for the density matrix in the spatial subspace. Similarly $\hat\beta_e$ denotes a basis in the electronic subspace.

Optical pumping is described by a linear first order differential equation for the density matrix $\hat\rho$, in the form

\begin{equation}
i \partial_t \hat\rho = \mathcal{L}\hat\rho,
\end{equation}
where $\mathcal{L}$ is a linear operator on $\hat\rho$.
This linear equation can be solved by matrix exponentiation:

\begin{equation}
\rho(t) = e^{-i \mathcal{L} t}\hat\rho(0)=U(t)\hat\rho(0),
\end{equation}
where the evolution operator $U(t)=e^{-i \mathcal{L} t}$ is the matrix exponentiation of $\mathcal{L}$.

Given the basis of $\hat\rho$, we can expand the linear operator $U$:

\begin{equation}
\begin{split}
    U&=\sum\limits_{\alpha_1,\alpha_2,\beta_1,\beta_2}A_{\alpha_1,\alpha_2,\beta_1,\beta_2}(\hat\alpha_{1,x}\hat\alpha_{2,x})\otimes(\hat\beta_{1,e}\hat\beta_{2,e})\\
    &=\sum\limits_{u,v}A_{u,v}\hat u_x \otimes\hat v_e.
\end{split}
\end{equation}
Here $\hat u_x$ and $\hat v_e$ again denote basis in the spatial and electronic subspace.

The excitation fraction is the probability of the atom to be found in a 'pumped' state $|p\rangle$ in the electronic subspace, and is given by

\begin{equation}
    \mathcal{F} = \mathrm{Tr}(\hat{P}\hat\rho),
\end{equation}
where $\hat{P} = |p\rangle\langle p|$ is the projection operator onto $|p\rangle$.
Therefore the excitation fraction after evolution $U$ from a initial pure state $\hat\rho_0=\hat\rho_{x,0}\otimes\hat\rho_{e,0}$ is

\begin{equation}
\begin{split}
    \mathcal{F} &= \mathrm{Tr}(\hat{P}U\hat\rho_0)\\
    &=\mathrm{Tr}(\sum\limits_{u,v} A_{u,v}\hat u_x\hat\rho_{x,0}\otimes \hat{P}\hat v_e\rho_{e,0})\\
    &=\sum\limits_{u,v} A_{u,v}\mathrm{Tr}(\hat u_x\hat\rho_{x,0})\mathrm{Tr}(\hat{P}\hat v_e\rho_{e,0})\\
    &=\mathrm{Tr}\left(\left[\sum\limits_{u,v} A_{u,v} \mathrm{Tr}(\hat{P}\hat v_e\rho_{e,0})\hat u_x\right]\rho_{x,0}\right)\\
    &=\mathrm{Tr}\left(f\rho_{x,0}\right),
    \end{split}
\end{equation}
where we denote $f = \sum\limits_{u,v} A_{u,v} \mathrm{Tr}(\hat{P}\hat v_e\rho_{e,0})\hat u_x$. This $f$ is in fact the excitation fraction of atoms in a uniform drive field, as in that case spatial degree of freedom is decoupled and $\mathcal{F}=\mathrm{Tr}\left(f\rho_{x,0}\right) = f$. Noting that the action of $f$ on $\rho_{x,0}$ is direct multiplication, we have 

\begin{equation}
    \mathcal{F}(\Delta x)=\int n(x) f(\Delta x-x) dx,
\end{equation}
where $n(x)$ is the diagonal of $\rho_{x,0}$ and is the initial spatial distribution, and $f(\Delta x - x)$ is the local excitation fraction due to an optical pumping lattice displaced by $\Delta x$.
As the remaining fraction $g \equiv 1-f$ and $\int n(x) dx=1$, we derive Eq.~\ref{eqNonLin} in the main text.
\begin{equation}
\begin{split}
    1-\mathcal{F}(\Delta x)&=1-\int n(x)\left[1-g(\Delta x-x)\right] dx\\
    &=\int n(x)g(\Delta x-x) dx.
\end{split}
\end{equation}

\subsection{Three-state optical pumping model}

Here we derive Eqs.~\ref{eqConv} and \ref{eqresolution} in the main text through quantitatively describing the optical pumping process.

We consider the states $F = 3,4,4'$. Spontaneous emission of $F'=4$ state does not always result in state $F=4$, but also state $F=3$.
The branching ratio for the desired decay into $F=4$ is $\beta = 7/12$.
In addition, we always operate in the regime where the pulse duration is much longer than the natural lifetime $1/\Gamma$ of the $F' = 4$ state, such that Rabi oscillations can be neglected.
Therefore we employ a three state rate equation model:
\begin{equation}\label{eqratesupp}
\begin{split}
    \dot{p_{4'}}&=-\frac{s\Gamma}{2}(p_{4'}-p_3)-\Gamma p_{4'}\\
    \dot{p_{3}}&=-\frac{s\Gamma}{2}(p_3-p_{4'})+(1-\beta)\Gamma p_{4'}\\
    \dot{p_4}&=\beta \Gamma p_{4'}\\
    f &= p_4,
\end{split}
\end{equation}
where $p$ denote occupation probabilities for different internal states, the excitation fraction $f$ is equal to the probability $p_4$ of the atom to be in $F=4$ state, and $s=2\Omega^2/\Gamma^2=I/I_\mathrm{sat}$ is the intensity in units of saturation intensity. 

Such a first order linear differential equation can be easily solved by matrix diagonalization.
The solution of $f$ at pulse time $t$ and intensity $s$ corresponding to drive field $\Omega$ is found to be: 

\begin{equation}\label{eqfracsupp}
\begin{split}
    f&=1-\frac{\gamma_+}{\gamma_+-\gamma_-}e^{-\gamma_-t}-\frac{\gamma_-}{\gamma_--\gamma_+}e^{-\gamma_+ t}\\
    \gamma_{\pm}&=\frac{\Gamma}{2}(s+1)\left(1\pm\sqrt{1-2s\beta/(s+1)^2}\right).
\end{split}
\end{equation}

The optical pumping lattice formed by retro-reflecting a beam with intensity $I$ gives rise to a drive field described by:
\begin{equation}
    \Omega(x)= \sqrt{2 s_0\Gamma^2} \sin (2\pi x/\lambda_\mathrm{op}),
\end{equation}
where $s_0 = I/I_\mathrm{sat}$ and $\lambda_\mathrm{op} = 852.335$ nm is the wavelength of the optical pumping light.
In the limit of long pulse time $t\gg 1/\Gamma$ where we operate, we can consider only the case with $s\ll 1$, as elsewhere $f \approx 1$. In this case,

\begin{equation}
    f = 1-e^{-\frac{\beta \Gamma}{2}\frac{s}{s+1}t},
\end{equation}
where $s = 2\Omega^2/\Gamma^2 = 4 s_0 \sin^2(2\pi x/\lambda_\mathrm{op})$. Therefore 

\begin{equation}
    g = e^{-\frac{\beta \Gamma}{2}\frac{s}{s+1}t}.
\end{equation}

The full width at half maximum $w$ of $g(x)$ is given by equation $g(w/2)=g(0)/2$. In the regime of super-resolution where $w \ll \lambda_\mathrm{op}$, the solution is

\begin{equation}\label{eqresolutionsupp}
    w = \frac{\lambda_\mathrm{op}}{2\pi}\sqrt{\frac{2\ln{2}}{s_0 t \beta\Gamma}}.
\end{equation}

This describes the predicted resolving power and its scaling with pumping power and pulse time in the strong pulse, long time limit.

The theoretical resolution shown in Fig 2 c in the main text is obtained differently, without making analytical approximations. Instead, the shown prediction is the FWHM of a numerically fitted Gaussian to the shape $f(x)$, the same way FWHM is extracted from experimental data.

\subsection{Numerical simulation of motional dynamics}

\begin{figure*}[htp]
\begin{center}
\includegraphics[width=170mm]{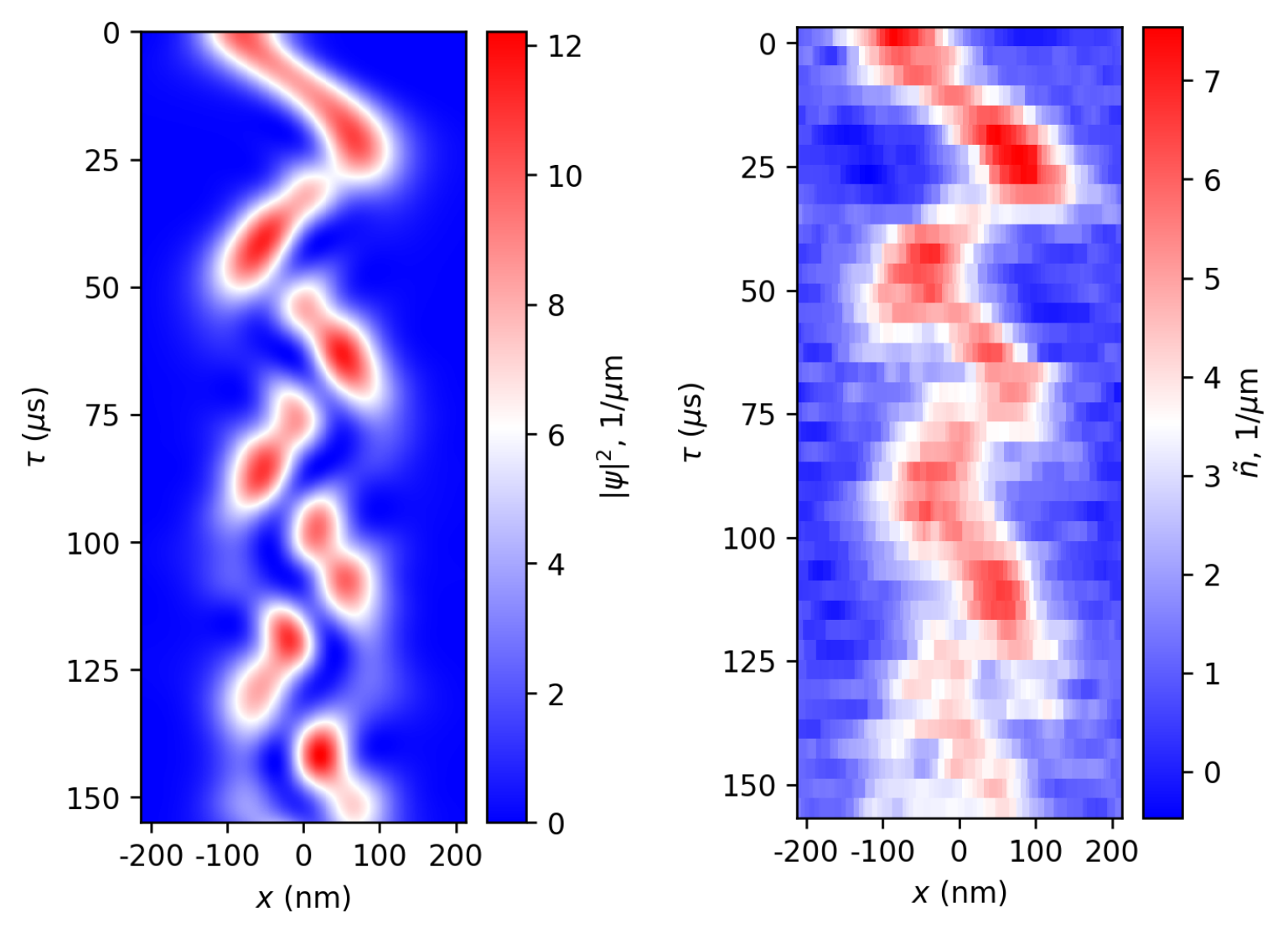}
\caption{Dynamics of an atom in a lattice site. Left: numerical simulation. Right: experimental data from Fig.~3b.}
\label{SFigdynamics}
\end{center}
\end{figure*}

Dynamics of a single particle in a sinusoidal optical lattice is given by the Schroedinger equation:

\begin{equation}
    i \hbar \partial_t \psi = -\frac{\hbar^2}{2m}\partial_x^2\psi-V_0\cos(4\pi x/\lambda)\psi.
\end{equation}

Given an initial condition $\psi_0$, this equation can be numerically solved by Fourier transform followed by matrix exponentiation, or projecting onto the basis of Mathieu functions, which are eigenstates of the Hamiltonian. We simulated the dynamics in a lattice with trap frequency 24~kHz, of an initial state that is the ground band Wannier function localized in one lattice site which is then shifted by 79~nm. The resulting $|\psi(\tau)|^2$ is plotted against $\tau$ in the left part of Fig.~S2. Comparing with measured data in Fig 3b, shown in the right part, simulation results reflect various features observed experimentally, including the non-sinusoidal motion of the peak, and the distortion of the wavefunction at later times. The simulation showed negligible tunneling to adjacent sites at 160~$\mu$s. Inhomogeneity of the traps along imaging direction is not included in simulation, and its contribution to damping of the observed dynamics cannot be reflected by the simulation.

\subsection{Imaging resolution}

This section will describe several systematic sources of broadening in the super-resolution signal.

\subsubsection*{Probe width}

The finite width of the super-resolution probe is determined by experimental parameters as described in the previous sections. The numerically predicted lineshape for a 1.4~$\mu$s pulse can be fitted with a Gaussian and plotted against $I/I\mathrm{sat}$ as shown in Figure 2c. For sufficiently high OP intensity, the width becomes smaller than that of the atomic density distribution.

\subsubsection*{Width of simple harmonic oscillator (SHO) thermal state}

The absolute ground state of a simple harmonic oscillator (SHO) has a probability distribution with $1\sigma$ width given by $\sigma_0=\sqrt{\hbar/2m\omega}$ where $m$ is the mass of caesium and $\omega$ is the trap frequency. Since the atoms in our sample are not all in the ground state, but rather are in a thermal ensemble comprising of the ground state and excited states, the actual width will be broadened. This can be taken into account by computing the thermal state probability density distribution $P(x)$ via a Boltzmann-weighted sum of the SHO eigenstates. With distance in units of the harmonic oscillator length $\sqrt{\hbar/m\omega}$, the probability distribution of an atom in a thermal ensemble at temperature $T$ can be written as:

\begin{equation}
P(x)=\frac{1}{Z}\sum_{n=0}^{\infty} \exp^{-E_n/k_B T}\left( \frac{1}{2^n n!}\pi^{-1/2}e^{-x^2}H_n(x)^2\right),
\end{equation}

\noindent where $Z=\left[ 2\sinh\left(\frac{hf}{2k_BT}\right)\right]^{-1}$ is the partition function, $E_n=\left(n+\frac{1}{2}\right)hf$ are the SHO energy eigenvalues, $k_B$ is Boltzmann’s constant, and $H_n(x)$ are the Hermite polynomials. Computing the sum shows that the distribution is Gaussian:


\begin{equation}
    P(x) \propto \exp \left[ -\left( 1-\frac{2e^{-hf/k_BT}}{1+e^{-hf/k_BT}} \right) x^2 \right].
\end{equation}

\noindent The $1\sigma$ width of the probability of an atom in a thermal ensemble with temperature $T$ is given by:

\begin{equation}
\sigma(T)=\sigma_0\sqrt{\coth{\left(\frac{hf}{2k_BT}\right)}}.
\end{equation}

\noindent Using results from a time-of-flight temperature measurement, we compute the predicted ground and thermal state widths to be 40(2) and 45(2)~nm, respectively (see Figure 2c).

\subsubsection*{Other systematic sources of blurring}

\paragraph{Misalignment between OP and trapping lattices.}

Our absorption imaging scheme involves column integration of a 3-dimensional atom cloud onto the CCD plane. Suppose the two lattices are misaligned by an angle $\alpha$, and the moir\'{e} pattern rotates by an angle $M\alpha$. Then, due to the different periodicities of the two lattices, the phase evolution across the cloud in the imaging direction $z$ is given by
\begin{equation}
    \phi(z)=\frac{\sin M\alpha}{M}z.
\end{equation}

\noindent Due to column integration, this effectively blurs the observed excitation fraction by a width

\begin{equation}
    \sigma_\phi=\frac{\sin M\alpha}{M}\sigma_\mathrm{at},
\end{equation}

\noindent where $\sigma_\mathrm{at}$ is the width of the atom cloud. 

\paragraph{Binning.}
Since the trapping and OP lattices have a frequency difference $\delta$, there is a linear phase gradient as shown in Figure 4b. The relative phase between the two lattices can be expressed (in units of length) as $\Delta\phi=\frac{\Delta f}{f}x=x/M$, where $x$ is the position along the lattice direction and $M$ is the same magnification as in Eq.~\ref{eqMag}. Since we wish to resolve spatial features of the atomic density, it is important to choose a bin width that is much smaller than $M \sigma(T)$. For the data presented in Figure 2 with $\delta=10$~GHz, a 10-pixel wide bin samples over $\Delta\phi=1.7$~nm, which is negligible compared to atomic feature sizes.

\end{document}